\documentstyle[12pt]{article}

\begin{document}

\begin{titlepage}

\begin{flushright}
TECHNION-PH-97-05\\
IJS-TP-97/07\\

%April 97\\ 
\end{flushright}

\vspace{.5cm}

\begin{center}
{\Large \bf Long distance $c \to u \gamma$ effects in weak radiative decays 
of D-mesons\\}

\vspace{1.5cm}

{\large \bf S. Fajfer$^{a,b}$ and
P. Singer$^{a}$\\}

\vspace{.5cm}

{\it a) Department of Physics, Technion - Israel Institute  of Technology, 
Haifa 32000, Israel\\}

\vspace{1cm}
{\it b) J. Stefan Institute, Jamova 39, P. O. Box 300, 1001 Ljubljana, 
Slovenia}
\end{center}

\centerline{\large \bf ABSTRACT}

\vspace{0.5cm}
We present a detailed analysis of the $D \to V \gamma $ 
transitions, using a model which combines 
heavy quark effective theory and the 
chiral Lagrangian approach and includes symmetry breaking. 
We notice that in addition to the previously considered 
s - channel annihilation and t - channel W - exchange, there is a 
long distance penguin - like $ c \to u \gamma$ contribution in the t - channel 
of Cabibbo - suppressed modes. Its magnitude is determined 
by the size of symmetry breaking which we calculate with a vector dominance 
approach. Although smaller in magnitude, the penguin - like 
contribution would lead to sizeable effects in case of cancellations  
among the other contributions to the 
amplitude. Thus, it may invalidate suggested tests for beyond the standard model 
effects in these decays. We also indicate the range of expectations for the 
branching ratios of various $D \to V \gamma$ modes. \\

\end{titlepage}

\setlength {\baselineskip}{0.75truecm}
\setcounter{footnote}{1}    % start footnotes at dagger instead
			     % of '*' (article style only)

\setcounter{footnote}{0}

\vskip 1cm

{\bf I. INTRODUCTION}\\

The study of weak radiative decays of charmed mesons is still 
in its early developing stage. No such 
events have been observed so far and there is no 
published upper limits yet \cite{PDG} for weak decays of $D$ 
mesons involving emission of real photons. On the theoretical 
front, the treatment of these decays 
faces a different situation than encountered in the amply studied 
weak radiative decays of B and K mesons. In the former, flavour - changing 
radiative decays can be interpreted at the quark level as dominated by 
the short - distance electromagnetic penguin, i.e. 
the $b \to s \gamma$ transition \cite{Desh}. In K - meson decays, both 
short - distance and long -  distance contributions may 
compete in various transitions \cite{batt}. 
In the weak radiative decays of charmed particles, the short - distance 
$c \to u \gamma$ process has been shown
to give a negligible contribution 
\cite{BGHP,GHMW}; as a result, these decays present 
the challenge and opportunity of 
developing the required theoretical treatment for the long - distance  
dynamics involved. 

The importance of a reliable description for these 
long - distance transitios is enhanced by the 
observation \cite{BGM,BFO0} that these decays 
provide also the interesting possibility for testing physics 
beyond the Standard Model, particularly non - minimal 
supersymmetric models. 

The first comprehensive phenomenological analysis 
of all possible $D \to V \gamma$ weak decays has been 
presented only recently \cite{BGHP}; several other papers 
have considered \cite{BFO0,BFO1,AK,HYC} specific modes,  
using various models. On the basis of the Standard Model 
weak Hamiltonian with QCD 
corrections \cite{bauertwo}, it emerges 
\cite{BFO0,BFO1} that in the quark picture without 
symmetry breaking the neutral charmed meson radiative decays can be viewed  
as due to W - exchange in the $u (t)$ channel, their amplitude being 
proportional to the $a_2$ Wilson coefficient. The charged 
decays on the other hand, evolve from annihilation diagrams 
and their amplitude 
is proportional to the $a_1$ Wilson coefficient. 

In the present paper we calculate the decay amplitudes for nine $D \to V \gamma$ 
transitions, by using a hybrid model which is a combination of 
heavy quark effective theory (HQET) and chiral Lagrangian \cite{BFO1}. 
We include systematically $SU(3)$ breaking into the amplitudes derived 
from it with particular attention payed to the coupling of vector mesons 
to photons as determined from experiment. 

An additional contribution to the radiative decays is due to the long - 
distance $c \to u \gamma$ transition. The similar contributions 
involved in the radiative decays of 
b and s quarks have been analyzed recently in several papers. 
The basic idea is to consider \cite{TD} the long - distance penguin - like 
quark process $Q \to q + V$, in which the pairs of quark - antiquark 
produced in the weak process materialize into vector mesons. These 
vector mesons are then allowed to convert into photons via the 
usual vector dominance process. Using 
this procedure, the long distance effects in $B  \to s (d) \gamma$ 
have been estimated \cite{TD,EIMS} with improved accuracy, as well as the 
effect in the $B \to X l^+ l^-$ decays \cite{A}. 
In Ref. \cite{EIMS} it was shown that the 
long - distance $s \to d \gamma$ transition is likely to be significantly 
larger than the short distance one. It was also 
pointed out in this paper that the size of this long distance contribution, 
estimated with vector meson dominance is determined by flavour symmetry 
breaking. \\

In this paper we consider the effect of the "long - distance penguin" 
embodied in the $c \to u \gamma$ transition, in the $D$ mesons weak radiative decays. 
(The role of this transition in charmed baryon decays is discussed in Ref. 
\cite{PZ}). 
When we include this contribution which is proportional to the $a_2$ Wilson 
coefficient, with $SU(3)$ symmetry breaking, we find that this newly considered contribution is present in 
the Cabibbo - suppressed weak radiative decays of charm mesons. 
The inclusion of the long distance $c \to u \gamma$ affects certain 
simple relations among the decay amplitudes 
$(D^0\to \rho^0 \gamma)/ (D^0\to \bar K^{*0} \gamma)$ and 
$(D^+_s \to K^{*+}\gamma)/D^+_s\to \rho^+ \gamma)$ which were noted 
previously \cite{BGM,BFO0} and were suggested as 
possible tests for new physics. 
The effect is not present in the amplitudes of the Cabibbo allowed decays 
$D^0 \to \bar K^{*0} \gamma$ and $D_s^+ \to \rho^+ \gamma$. 
We point out that as a result of the GIM mechanism, 
the long - distance 
$c \to u \gamma$ contribution will vanish in case of exact $SU(3)$ 
symmetry.

Using a hybrid model \cite{BFO1} and including systematically $SU(3)$ breaking 
into amplitudes derived from it, with particular attention 
paid to the coupling of vector mesons to photons as 
derived from experiment, 
we calculate all the $D \to V \gamma$ transitions. 
The numerical values of these amplitudes are displayed in Tables 2 and 3. 
Since the relative phase of different contributions is unknown, we 
cannot make firm 
predictions for the expected rates. Nevertheless, 
taking this uncertainty into account and after fixing some of the constants of 
the model, we are able to indicate a fairly limited expected range 
for branching ratios of certain modes. 
Thus, we show that the Cabibbo allowed 
$D_s^+ \to \rho ^+ \gamma$ is expected to have 
a branching ratio of $(3 - 4.5) \times 10^{-4}$, 
while the Cabibbo - suppressed decays $D_s^+ \to K^{*+} \gamma$, 
$D^+ \to \rho^+ \gamma$ are expected to occur with branching ratios 
in the $(2 - 4) \times 10^{-5}$ range. \\

In Section II we present the theoretical framework for our calculation. 
In Section III we display the explicit expressions of all 
calculated decay amplitudes and in Section IV we summarize 
and compare with previous calculations. \\

{\bf II. THE THEORETICAL FRAMEWORK}\\

In this section we present in detail the 
theoretical basis needed for the calculation 
of the $D \to V \gamma$ amplitudes, which evolve 
from long - distance dynamics. This basis covers the strong 
and weak interactions sectors and along 
the presentation we explain our considerations for the choice 
of relevant numerical parameters. \\

{\bf A. Chiral lagrangians, heavy quark limit and vector meson \\
dominance}\\

We incorporate in our Lagrangian \cite{BFO2} 
both the heavy flavor $SU(2)$ symmetry \cite{IW},  \cite{georgi} and 
the $SU(3)_L\times SU(3)_R$ chiral 
symmetry, spontaneously broken to the diagonal 
$SU(3)_V$ \cite{bando}, which can be used for the 
description of heavy and light pseudoscalar and 
vector mesons. A similar Lagrangian, but without the 
light vector octet, was first introduced by Wise 
\cite{W}, Burdman and Donoghue \cite{BURDMAN}, 
and Yan et al. \cite{YAN}. 
It was then generalized with the inclusion of light 
vector mesons by Casalbuoni et al. \cite{casone}.

The light degrees of freedom are described by the 
3$\times$3 Hermitian matrices 

\begin{eqnarray}
\label{defpi}
\Pi = \pmatrix{
{\pi^0 \over \sqrt{2}} + {\eta_8 \over \sqrt{6}}+ {\eta_0 \over \sqrt{3}}& \pi^+ & K^+ \cr
\pi^- & {-\pi^0 \over \sqrt{2}} + {\eta_8 \over \sqrt{6}} + 
{\eta_0 \over \sqrt{3}} & K^0 \cr
K^- & {\bar K^0} & -{2 \eta_8 \over \sqrt{6}} + {\eta_0 \over \sqrt{3}} \cr}
\end{eqnarray}

\noindent
and

\begin{eqnarray}
\label{defrho}
\rho_\mu = \pmatrix{
{\rho^0_\mu + \omega_\mu \over \sqrt{2}} & \rho^+_\mu & K^{*+}_\mu \cr
\rho^-_\mu & {-\rho^0_\mu + \omega_\mu \over \sqrt{2}} & K^{*0}_\mu \cr
K^{*-}_\mu & {\bar K^{*0}}_\mu & \Phi_\mu \cr}
\end{eqnarray}

\noindent
for the pseudoscalar and vector mesons, respectively. They are 
usually expressed through the combinations 

\begin{eqnarray}
\label{defu}
u & = & \exp  ( \frac{i \Pi}{f} )\;,
\end{eqnarray}

\noindent 
where $f\simeq f_{\pi}=132$ MeV is the pion pseudoscalar decay constant, and 

\begin{eqnarray}
\label{defrhohat}
{\hat \rho}_\mu & = & i {{\tilde g}_V \over \sqrt{2}} \rho_\mu\;,
\end{eqnarray}

\noindent
where ${\tilde g}_V=5.9$ was fixed in the case of exact flavor symmetry \cite{bando}. 
In the following we will also use gauge field tensor $F_{\mu \nu}({\hat \rho})$

\begin{eqnarray}
F_{\mu \nu} ({\hat \rho}) = 
\partial_\mu {\hat \rho}_\nu -
\partial_\nu {\hat \rho}_\mu +
[{\hat \rho}_\mu,{\hat \rho}_\nu]. 
\label{deff}
\end{eqnarray} 
It is convenient to introduce two currents
${\cal V}_{\mu} =  \frac{1}{2} (u^{\dag}
D_{\mu} u + u D_{\mu}u^{\dag})$
and ${\cal A}_{\mu}  =  \frac{1}{2} (u^{\dag}D_{\mu} u - u D_{\mu}u^{\dag})$. 
The covariant derivative of $u $ and $u^{\dag}$ is 
defined as $D_{\mu}u = (\partial_{\mu} + {\hat B}_{\mu} )u$ 
and $D_{\mu}u^{\dag}= (\partial_{\mu} + {\hat B}_{\mu} )u^{\dag}$,  
with ${\hat B}_{\mu} = i e B_{\mu} Q$, 
$Q = diag (2/3,-1/3,-1/3)$, $B_{\mu}$ being the photon field.

The light meson part of the strong 
Lagrangian can be written as \cite{bando} 
\begin{eqnarray}
\label{deflight}
{\cal L}_{light} & = & -{f^2 \over 2}
\{tr({\cal A}_\mu {\cal A}^\mu) +
a\, tr[({\cal V}_\mu - {\hat \rho}_\mu)^2]\}\nonumber\\
& + & {1 \over 2 {\tilde g}_V^2} tr[F_{\mu \nu}({\hat \rho})
F^{\mu \nu}({\hat \rho})]\;.
\end{eqnarray}
The constant $a$ in (\ref{deflight}) is in principle a free parameter. 
In the case of exact vector meson dominance (VDM) $a=2$ \cite{bando,fuj}. 
However, the photoproduction and decays data indicate \cite{EIMS}  
that the $SU(3)$ breaking modifies the VDM in 
\begin{eqnarray} 
\label{VDM}
{\cal L}_{V-\gamma} = -e {\tilde g}_V f^2 B_{\mu} (\rho^{0\mu} + \frac{1}{3} 
\omega^{\mu} - \frac{{\sqrt 2}}{3} \Phi^{\mu}).
\end{eqnarray}
Instead of the exact $SU(3)$ limit (${\tilde g}_V= m_V/f$),  
we shall use  the measured values, defining 
\begin{eqnarray} 
< V(\epsilon_V,q)| V_{\mu} |0> = i \epsilon_{\mu}^{*} (q) g_V(q^2). 
\label{defg}
\end{eqnarray}
The couplings $g_V(m_V^2)$ are obtained from the leptonic decays of these mesons. 
In our calculation we use 
$g_{\rho}(m_{\rho}^2) \simeq g_{\rho}(0)= 0.17 GeV^2$, 
$g_{\omega}(m^2_{\omega}) \simeq g_{\omega} (0) = 0.15 GeV^2$ 
and $g_{\Phi}(m^2_{\Phi}) \simeq g_{\Phi} (0) = 0.24 GeV^2$.

Both the heavy pseudoscalar and the heavy vector 
mesons are incorporated in a $4\times 4$ matrix 
\begin{eqnarray}
\label{defh}
H_a& = & \frac{1}{2} (1 + \!\!\not{\! v}) (P_{a\mu}^{*}
\gamma^{\mu} - P_{a} \gamma_{5})\;,
\end{eqnarray}
\noindent
where $a=1,2,3$ is the $SU(3)_V$ index of the light 
flavors, and $P_{a\mu}^*$, $P_{a}$, annihilate a 
spin $1$ and spin $0$ heavy meson $Q \bar{q}_a$ of 
velocity $v$, respectively. They have a mass dimension 
$3/2$ instead of the usual $1$, so that the Lagrangian 
is in the heavy quark limit $m_Q\to\infty$ explicitly 
mass independent. Defining moreover 
\begin{eqnarray}
\label{defhbar}
{\bar H}_{a} & = & \gamma^{0} H_{a}^{\dag} \gamma^{0} =
(P_{a\mu}^{* \dag} \gamma^{\mu} + P_{a}^{\dag} \gamma_{5})
\frac{1}{2} (1 + \!\!\not{\! v})\;,
\end{eqnarray}

\noindent
we can write the strong  Lagrangian as 
\begin{eqnarray}
\label{deflstrong}
{\cal L}_{even} & = & {\cal L}_{light} +
i Tr (H_{a} v_{\mu} D^{\mu} 
{\bar H}_{a})\nonumber\\
& + &i g Tr [H_{b} \gamma_{\mu} \gamma_{5}
({\cal A}^{\mu})_{ba} {\bar H}_{a}]
 +  i \beta Tr [H_{b} v_{\mu} ({\cal V}^{\mu}
- {\hat \rho}_{\mu})_{ba} {\bar H}_{a}]\nonumber\\
& + &  {\beta^2 \over 4 f^2 }
Tr ({\bar H}_b H_a {\bar H}_a H_b)\;.
\end{eqnarray}
where
$ D^{\mu} {\bar H}_{a} = (\partial_{\mu} + {\cal V}_{\mu} - 
i e Q^{\prime} B_{\mu}){\bar H}_{a}$, with $Q^{\prime} = 2/3$ 
for c  quark \cite{BFO1}. 

It contains two unknown parameters, $g$ and $\beta$, which 
cannot be determined by symmetry arguments, but must 
be fitted by experiment. It is the most general 
even-parity Lagrangian in the leading heavy quark mass 
($m_Q\to\infty$) and chiral symmetry ($m_q\to 0$ 
and the minimal number of derivatives) limit. 

The electromagnetic field can couple to the mesons also through the anomalous 
interaction; i.e., through the odd parity Lagrangian. 
The contributions to this Lagrangian arise from terms  of the Wess - Zumino - Witten 
kind,  given by \cite{fuj,BGP,H}

\begin{eqnarray}
\label{defloddlight}
{\cal L}^{(1)}_{odd} & = & -4 \frac{C_{VV\Pi}}{f} \epsilon
^{\mu \nu \alpha \beta}Tr (\partial_{\mu}
{\rho}_{\nu} \partial_{\alpha}{\rho}_{\beta} \Pi). 
\end{eqnarray}

The coupling $C_{VV\Pi}$ can be  
determined in the case of the exact $SU(3)$ flavor symmetry 
following the hidden symmetry approach of \cite{bando,fuj} and 
it is found to be $C_{VV\Pi} = 3{\tilde g}_V^2 /32 \pi^2 = 0.33$.  
In the following we use the VDM 
(\ref{VDM}), however we allow for $SU(3)$ symmetry breaking 
in the couplings of vector mesons to photon, which is expressed 
by the physical values of $g_V$ and $m_V$. 
The decay width for $V \to P\gamma$ can be written as 
\begin{equation}
\Gamma (V \to P \gamma) = \frac{\alpha}{24} \frac{(m_V^2 - m_P^2)^3}{m_V^3}
|g_{VP\gamma}|^2 .
\label{decvpg}
\end{equation}

Using (\ref{defloddlight}) and (\ref{VDM})
modified as explained above, and taking the experimental value 
for the K - meson decay coupling $f_K =  0.160$ $GeV$ we find 

\begin{equation}
g_{\omega \pi \gamma} = 4 \frac{g_{\rho}}{m_{\rho}^2}
\frac{C_{VV\Pi}}{f_{\pi}},
\label{decom}
\end{equation}
\begin{equation}
g_{\rho \pi \gamma} = 4 \frac{g_{\omega}}{3 m_{\omega}^2}
\frac{C_{VV\Pi}}{f_{\pi}},
\label{decrho}
\end{equation}
\begin{equation}
g_{K^{*+} K^+ \gamma} =2 ( \frac{g_{\omega}}{3 m_{\omega}^2} + 
\frac{g_{\rho}}{m_{\rho}^2} - \frac{2}{3} \frac{g_{\Phi}}{m_{\Phi}^2})
\frac{C_{VV\Pi}}{f_K},
\label{deckp}
\end{equation}
\begin{equation}
g_{K^{*0} K^0 \gamma} =2 ( \frac{g_{\omega}}{3 m_{\omega}^2 }- 
\frac{g_{\rho}}{m_{\rho}^2 }- \frac{2}{3} \frac{g_{\Phi}}{m_{\Phi}^2})
\frac{C_{VV\Pi}}{f_K}.
\label{decko}
\end{equation}

At this point, we have the choice of using the 
symmetry value $C_{VV\Pi} = 0.33$, or using a best fit by comparing (\ref{decom}) 
- (\ref{decko}) to the experimental values \cite{PDG}: 
$\Gamma (\omega \to \pi \gamma) = (7.25 \pm 0.5) \times 10^{-4}$ $GeV$, 
$\Gamma (\rho^+ \to \pi^+ \gamma) = (6.8 \pm 0.6)\times 10^{-5}$ $GeV$, 
$\Gamma (\rho^0 \to \pi^0 \gamma) = (1.2 \pm 0.3)\times 10^{-4}$ $GeV$, 
$\Gamma ( K^{*+} \to K^+ \gamma) = (5.0 \pm 0.5)\times 10^{-5}$ $GeV$, 
$\Gamma (K^{*0} \to K^0 \ \gamma) = (1.2 \pm 0.1)\times 10^{-4}$ $GeV$. 
We choose as a best fit the value $C_{VV\Pi} = 0.31$, which reproduces the 
observed width of $K^{*+} \to K^+ \gamma$ and gives 
$\Gamma (\omega \to \pi \gamma) = 9.8 \times 10^{-4}$ $GeV$,  
$\Gamma (\rho \to \pi \gamma) = 7.7 \times  10^{-5}$ $GeV$, 
$\Gamma (K^{*0} \to K^0 \ \gamma) = 1.42\times 10^{-4}$ $GeV$. 
Comparing these figures with the experimental results, it is obvious that the 
inclusion of VDM with $SU(3)$ breaking improves the results obtained 
\cite{fuj} for exact $SU(3)$. 
We remark that the inclusion of 
$SU(3)$ symmetry breaking effects for these decays has been suggested 
often, including its inclusion in $C_{VV\Pi}$ \cite{BGP} - \cite{BMS}. 
Our approach here takes into account VDM with the observed values 
of $g_V$ and $m_V$, without additional symmetry breaking 
parameters, while 
the approaches of \cite{BGP,H} make fits using available experimental data on 
$\Gamma (V \to P \gamma)$. 

We will also need the odd - parity  Lagrangian in the heavy sector. 
There are two contributions \cite{BFO1,CFN} in it, 
characterized by coupling strenghts $\lambda$ and $\lambda^{\prime}$. 
The first is given by 
\begin{eqnarray}
\label{defoddheavy}
{\cal L}_{1} & = & i {\lambda} Tr [H_{a}\sigma_{\mu \nu}
F^{\mu \nu} (\hat \rho)_{ab} {\bar H_{b}}].
\end{eqnarray}
In this term the interactions of light vector mesons with heavy pseudoscalar 
or heavy vector mesons is described. The light vector meson 
can then couple to the 
photon by the standard VDM prescription. 
This term is of the order ${1}/{{\lambda}_{\chi}}$ with 
${\lambda}_{\chi}$ being the chiral perturbation theory scale \cite{CG}.

The second term gives the direct heavy quark - photon interaction and 
is generated by the Lagrangian
\begin{eqnarray}
\label{defoddheavy2}
{\cal L}_{2} & = & - {\lambda}^{\prime} Tr [H_{a}\sigma_{\mu \nu}
F^{\mu \nu} (B) {\bar H_{a}}].
\end{eqnarray}
The parameter ${\lambda}^{\prime} $ can be approximatelly 
related to the charm quark magnetic moment via 
${\lambda}^{\prime} \simeq 1/(6 m_c)$ 
\cite{IW,YAN,CG,amun} and it should be considered as a higher order 
term in $1/m_Q$ expansion \cite{CG,amun}. 

In order to gain information on these couplings we turn to an 
analysis of $D^{*0} \to D^0 \gamma$, 
$D^{*+}\to D^+ \gamma$ and $D_s^{*+} \to D^+_s \gamma$ decays. 
Experimentally, only the ratios 
$R_{\gamma}^0 = \Gamma (D^{*0} \to D^0 \gamma)/\Gamma 
(D^{*0} \to D^0 \pi^0)$ and 
$R_{\gamma}^+ = \Gamma (D^{*+} \to D^+ \gamma)/\Gamma 
(D^{*+} \to D^+ \pi^0)$ are known \cite{PDG}. 
Using VDM with  (\ref{deflstrong}),(\ref{defoddheavy}), (\ref{defoddheavy2}) 
we calculate
\begin{equation}
R_{\gamma}^0 = 64 \pi \alpha \frac{f^2}{g^2} 
(\frac{p_{\gamma}^+}{p_{\pi}^0})^3 [ \lambda^{\prime} + \lambda 
\frac{{\tilde g}_V}{2 {\sqrt 2}}(\frac{g_{\rho}}{m_{\rho}^2} 
+\frac{g_{\omega}}{3 m_{\omega}^2})]^{2}
\label{r0}
\end{equation}
and 
\begin{equation}
R_{\gamma}^+ = 64 \pi \alpha \frac{f^2}{g^2} 
(\frac{p_{\gamma}^+}{p_{\pi}^0})^3 [ \lambda^{\prime} - \lambda 
\frac{{\tilde g}_V}{2 {\sqrt 2}}(\frac{g_{\rho}}{m_{\rho}^2} 
-\frac{g_{\omega}}{3 m_{\omega}^2})]^2
\label{r+}
\end{equation}
Numerically, the inclusion of $SU(3)$ breaking changes  
$|\lambda^{\prime} +  \frac{2}{3} \lambda |$ in $R^0_{\gamma}$ to become 
$|\lambda^{\prime} + 0.77\lambda|$ and 
$|\lambda^{\prime} - \frac{1}{3} \lambda|$ in $R^+_{\gamma}$, becomes 
$|\lambda^{\prime} - 0.427 \lambda|$. 
Taking the $R_{\gamma}^0 = 0.616$ and $R_{\gamma}^+ = 0.036$ \cite{PDG}, 
we obtain two sets of solutions for $|\lambda^{\prime} /g|$ and 
$|\lambda/g|$. The first is $|\lambda /g| = 0.533$ $GeV^{-1}$, 
$|\lambda^{\prime} /g |= 0.411$ $GeV^{-1}$ 
and second is $|\lambda /g| = 0.839$  $GeV^{-1}$, 
$|\lambda^{\prime} /g |= 0.175$  $GeV^{-1}$.
In our calculation we have  the combinations  
$|\lambda^{\prime} + 0.77\lambda| = 0.821|g|$ $GeV^{-1}$ and 
$|\lambda^{\prime} - 0.427 \lambda| = 0.183 |g|$  $GeV^{-1}$. 
For  $D_s^{+*} D_s^{+} \gamma$ one derives the coupling 
$|\lambda^{\prime} - \lambda {{\tilde g}_V}/{{\sqrt 2}} {g_{\Phi}}/{3 m_{\Phi}^2}| = 
|\lambda^{\prime} - 0.32 \lambda|$. 
Unfortunately, $R^+_{\gamma}$ is poorly known, essentally within factor of $3$ 
\cite{PDG}, which could induce rather large errors in our 
determination of these constants. On the other hand, one should mention that the values 
we use for $R_{\gamma}^+$, $R_{\gamma}^0$ fit well the theoretical 
expectations for these ratios, as determined in rather different models 
\cite{CFN} - \cite{MS}.

In addition to strong and electromagnetic interaction, we have to 
specify the weak one. The nonleptonic weak Lagrangian on the quark level can be written as usual 
\cite{bauertwo}
\begin{eqnarray}
\label{deflfermisl}
{\cal L}_{SL}^{eff}(\Delta c=\Delta s=1)
& = & -{G_F \over \sqrt{2}} V_{uq_i}V_{cq_j}^*  
[ a_1 ({\bar u} q_i)^{\mu}_{V-A}
({\bar q_j}c )_{V-A,\mu} \nonumber\\
& + &  a_2 ({\bar u} c)_{V-A,\mu} 
({\bar q}_j q_i)_{V-A}^{\mu}], 
\end{eqnarray}

\noindent
where $V_{ij}$ are the CKM matrix elements, $G_F$ is the 
Fermi constant and $({\bar \Psi}_1\Psi_2)^\mu\equiv
{\bar \Psi}_1\gamma^\mu(1-\gamma^5)\Psi_2$. 
In our calculation we use $a_1 = 1.26$ and $a_2 = -0.55$ as 
found in \cite{bauertwo}.

At the hadronic level, the weak current transforms as $({\bar 3}_L,1_R)$ 
under chiral $SU(3)_L\times SU(3)_R$ being linear in the 
heavy meson fields $D^a$ and $D^{*a}_\mu$ and is taken as  \cite{BFO2}

\begin{eqnarray}
\label{jqbig}
{J_Q}_{a}^{\mu} = &\frac{1}{2}& i \alpha Tr [\gamma^{\mu}
(1 - \gamma_{5})H_{b}u_{ba}^{\dag}]\nonumber\\
&+& \alpha_{1}  Tr [\gamma_{5} H_{b} ({\hat \rho}^{\mu}
- {\cal V}^{\mu})_{bc} u_{ca}^{\dag}]\nonumber\\
&+&\alpha_{2} Tr[\gamma^{\mu}\gamma_{5} H_{b} v_{\alpha} 
({\hat \rho}^{\alpha}-{\cal V}^{\alpha})_{bc}u_{ca}^{\dag}]+...\;,
\end{eqnarray}

\noindent
where $\alpha=f_H\sqrt{m_H}$ \cite{W}, $\alpha_1$ was 
first introduced by Casalbuoni et al. \cite{casone,castwo}, while $\alpha_2$ 
was introduced in \cite{BFO2}.
It has to be included, 
since it is of the same order in the $1/m_Q$ and chiral 
expansion as the term proportional to $\alpha_1$ \cite{BFO2}. \\

{\bf B. The $c \to u \gamma$ long distance contribution}\\

In addition to the photon interaction discussed in previous 
subsection it was noticed 
\cite{TD,EIMS} that the $SU(3)$ breaking 
causes a long distance penguin - like contribution 
proportional to the $a_2$ Wilson coefficient as shown in Fig. 1.
Knowing that $V_{ud} V_{cd}^* = - V_{us} V_{cs}^*$ and using 
factorization and the VDM one derives \cite{EIMS,PZ} the 
effective Hamiltonian for $c \to u \gamma$ transition
\begin{eqnarray}
\label{cupen}
{\cal H} (c \to u V \to u \gamma) = 
\frac{G_F}{{\sqrt 2}} a_2 {\bar u} \gamma^{\mu}(1- \gamma_5) c 
[-\frac{1}{2} \frac{g_{\rho}^2(0)}{m_{\rho}^2}
+ \frac{1}{6} \frac{g_{\omega}^2(0)}{m_{\omega}^2}
+ \frac{1}{3} \frac{g_{\Phi}^2(0)}{m_{\Phi}^2}]
\epsilon_{\mu}^{*\gamma}.
\end{eqnarray}
The expression in (\ref{cupen}) is not in the gauge invariant 
form needed for the replacement $V_i \to \gamma$ (see also Ref. \cite{TD}.)  
However, gauge invariance will be imposed when obtaining the 
matrix elements $D \to V \gamma$.
In the case of exact $SU(3)$ 
symmetry the expression in square brackets vanishes 
as a result of GIM cancellations. 
This effect was found to be significantly larger than the 
short distance one in the $s \to d \gamma$ \cite{EIMS} 
and $c \to u \gamma$ \cite{PZ} cases. 
In the case of $D \to V \gamma$ decay one has to evaluate the 
matrix element of $<V(p_V,\epsilon)| 
{\bar u} \gamma^{\mu} (1 - \gamma^5) c| D(p)>$. The relevant 
matrix element is parametrized usually in 
$D \to V l \nu_l$ semileptonic decay as \cite{BGHP,bauertwo,BFO2,RB}

\begin{eqnarray}
\label{parhv}
<V(p_V,\epsilon_V)|(V-A)^\mu|D(p)>=
{2 V(q^2)\over m_D+m_V}
\epsilon^{\mu\nu\alpha\beta}\epsilon_{V\nu}^* p_\alpha
{p_V}_\beta \nonumber\\
+i \epsilon^*_V.q {2 m_V\over q^2}q_\mu ( A_3(q^2) - A_0(0)) 
+i(m_D+m_V)[\epsilon_{V\mu}^* A_1(q^2) \nonumber\\
-{\epsilon^*_V .q\over m_D+m_V}((p+p_V)_\mu
A_2(q^2)] \;,
\end{eqnarray}
where $q = p - p_V$. In order that these matrix elements 
be finite at $q^2 = 0$, the form factors satisfy the relation 
\cite{bauertwo}  

\begin{equation}
\label{relff}
A_3(q^2)-{m_H+m_V\over 2 m_V}A_1(q^2)+
{m_H-m_V\over 2 m_V}A_2(q^2)=0\;,
\end{equation}
and $A_3 (0) = A_0(0)$. Imposing gauge invariance, one derives
\begin{eqnarray}
\label{parhv1}
\epsilon_{\mu}^{*\gamma}(q) <V(p_V,\epsilon_V)|(V-A)^\mu|D(p)>=
- {2 V(0)\over m_D+m_V}
\epsilon^{\mu\nu\alpha\beta}\epsilon_{\mu}^{*\gamma}\epsilon_\nu^{*V} 
p_{\alpha} {p_V}_\beta \nonumber\\
+i(m_D+m_V) A_1(0) [ \epsilon^{*\gamma} \cdot \epsilon^{*V}
-\epsilon^{*\gamma} \cdot p_V \epsilon^{*V} \cdot q \frac{1}{p_V \cdot q}]. 
\end{eqnarray}
Now, in order to obtain $V(0)$, $A_1(0)$ appearing in (\ref{parhv1}) we rely 
on the knowledge of 
form factors $|V^{DV}(0)|$ and $|A_1^{DV}(0)|$, determined 
in the semileptonic decays. 
Experimentally these form factors were extracted from the 
$D^+ \to {\bar K}^{0*} l \nu_l$ and $D_s^+ \to \Phi l \nu_l$ decays, assuming 
the pole behavior \cite{PDG}. The hybrid model of 
\cite{casone,castwo}, described in the previous section, works 
well for small recoil momentum in 
semileptonic decays, or equivalently in the case of 
$q^2_{max}$. We follow their approach, since the experimental 
extrapolation of form factors at $q^2= 0$ assumes 
their pole behavior. 
They have found \cite{castwo}
\begin{eqnarray}
V^{DV}(q^2_{max}) & = &- \frac{{\tilde g}_V}{{\sqrt 2}} \lambda f_{D^{\prime *} }
\frac{m_D + m_V}{v\cdot q_{max} + m_{D^{\prime *}} - m_D}.
\label{VQ}
\end{eqnarray}
where $m_{D^{\prime *}}$ denotes the mass of 
the corresponding vector meson pole. 
The monopole assumption leads to 
\begin{eqnarray}
V^{DV}(0)&  = &- \frac{{\tilde g}_V}{{\sqrt 2}} \lambda f_{D^{\prime *}} 
\frac{(m_D + m_V) (m_{D^{\prime *}} + m_D - m_V)}{ m_{D^{\prime *}}^2}. 
\label{VQ0}
\end{eqnarray}
For the $A_1$ form factor  the authors of \cite{castwo} found
\begin{eqnarray}
\label{A1}
A_1^{DV}(q^2_{max} ) & = &- \frac{{\tilde g}_V}{{\sqrt 2}} 
2 \alpha_1 \frac{{\sqrt m_D}}{m_D + m_V}
\end{eqnarray}
and at $q^2 = 0$  
\begin{eqnarray}
\label{A10}
A_1^{DV}(0)&  = & - {\tilde g}_V {\sqrt 2} 
\alpha_1 \frac{ {\sqrt m_D}}{m_D + m_V}
[1 - \frac{(m_D - m_V)^2}{m_{D_{1^+}}}],
\end{eqnarray}
where $D_{1^+}$ is the mass of the ${\bar q} c$ $J^{P}= 1^+$ 
bound state. 
(We use the masses of ${\bar s}c$ and  ${\bar d}c$ bound states 
to be $2.53$ $ GeV$ 
 and $2.42$ $GeV$ as in \cite{castwo}.) 
In  \cite{PDG} there are listed data on form  factors at $q^2 = 0$ 
obtained from 
$D^+ \to {\bar K}^{*0} l \nu_l$ and $D_s^+ \to \Phi l \nu_l$ 
decays. From $D^+ \to {\bar K}^{*0} l \nu_l$ decay 
the form  factors are 
$|V^{DK^*} (0)| = 1.0 \pm 0.2$, $|A_1^{DK^*} (0)| = 0.55 \pm0.03$ and 
$|A_2^{DK^*}(0)| = 0.40 \pm 0.08$. 
From $D_s^+ \to \Phi l \nu_l$ decay data it was extracted \cite{PDG} 
$|V^{D_s \Phi} (0)| = 0.9 \pm 0.3$, $|A_1^{D_s \Phi} (0)| = 0.62 \pm 0.06$ and 
$|A_2^{D_s \Phi}(0)| = 1.0  \pm 0.3$.  
Using the values $|V(0)|$ from both decays and taking the average 
of the two values derived from (\ref{VQ0}) for $|\lambda|$, 
we obtain $|\lambda| = 0.479$ $GeV^{-1}$. 
This gives $|g |= 0.58$, found from $|\lambda/ g| = 0.839$ $GeV^{-1}$. 
The value $|\lambda/ g| = 0.533$  $GeV^{-1}$ leads 
to somewhat higher value of $g$ 
than expected \cite{PDG}. \\
Using again the average of two experimental 
$A_1(0)$ values, we obtain 
$|\alpha_1| = 0.171$ $ GeV^{1/2}$. \\

{\bf III. The decay amplitudes}\\

The amplitudes for $D \to V \gamma$ can be written 
in the gauge  - invariant form

\begin{eqnarray}
A(D(p)  \to V(\epsilon_V,p_V) \gamma(\epsilon_{\gamma},q)) & = & 
e \frac{G_F}{{\sqrt 2}} V_{uq_i} V_{cq_j}^* 
\{\epsilon_{\mu \nu \alpha \beta} q^{\mu} \epsilon_{\gamma}^{*\nu} p^{\alpha}
\epsilon_V^{*\beta} A_{PC} \nonumber\\
 +  i ( \epsilon_V^{*\beta} \cdot q\epsilon_{\gamma}^{*\nu} \cdot p_V &-&
p_V \cdot q \epsilon_V^{*\beta} \cdot \epsilon_{\gamma}^{*\nu}) A_{PV}\}.
\label{amplitude}
\end{eqnarray}
The $A_{PC}$ and $A_{PV}$ denote the parity conserving  
and parity violating parts of the amplitude \cite{BGHP}. 
The different contributions to the decay amplitude arising in our model 
are displayed in schematic form in Fig. 2. 
The photon can be first emitted from the D meson which becomes $D^*$, 
which then weakly decays into vector meson $V$. 
Their vertices are proportional 
to a combination of $\lambda^{\prime}$ and $\lambda$,  
calculated in the analysis of  $D^* \to D \gamma$ decay amplitudes. 
We denote this part of amplitude as 
$A_{PC}^{(I)}$. 
When calculating these decay amplitudes, %using $\lambda^{\prime}$ and $\lambda$  
we have to remark that the $1/m_c$ corrections coming from light-quark 
current, effectively included into the $\lambda^{\prime}$ 
parameter, are not necessarily the same as in the case of $D^* \to D \gamma$. 
This uncertainty unfortunately increases present theoretical and experimental 
uncertainty already present in the calculation of $D \to V \gamma$. 
The second contribution comes from the weak decay of $D$ meson,  firstly 
into off - shell 
light pseudoscalar, which then decays into $V \gamma$. 
We denote this part of amplitudes as
$A_{PC}^{(II)}$. 

The charged charm meson can radiate a real photon from the term  
$-e v^{\mu} B_{\mu}$ $Tr [H_a (Q - 2/3 )_{ab} {\bar H}_b]$ 
given in (\ref{deflstrong}),  while the charged light vector meson can radiate  
through the last term of (\ref{deflight}). 
Both contributions are present in  $A_{PV}$. 
The $SU(3)$ breaking effects are accounted in $f_{D}$, $f_{D_s}$, $g_{K^*}$  and 
$g_{\rho}$ (we take $g_{K^*} = (m_{K^*}/m_{\rho})g_{\rho} $ ). 

The long distance penguin - like contribution is present 
in the Cabibbo suppressed charm meson decays. It contributes  to 
both the parity - conserving and the parity - violating 
parts of the decay amplitude. We denote its contribution as 
$A_{PC}^{(III)}$  and $A_{PV}^{(III)}$.  

The Cabibbo allowed decay amplitudes are proportional to 
the product $|V_{ud} V_{cs}^*|$. 
\begin{eqnarray}
A_{PC} (D^{0} \to {\bar K}^{*0} \gamma)&  = &4 a_2 | \lambda^{\prime} + \lambda 
\frac{{\tilde g}_V}{2{\sqrt 2}} (\frac{g_{\rho}}{m_{\rho}^2}
 + \frac{g_{\omega}}{3 m_{\omega}^2}) |
\frac{f_D g_{K^{*}} m_{D^*}}{m_{D^*}^2 - m_{K^*}^2} 
{\sqrt \frac{m_{D^*}}{m_D}}\nonumber\\
& + & 2 a_2 |C_{VV\Pi}|(\frac{g_{\omega}}{3 m_{\omega}^2} - 
\frac{g_{\rho}}{m_{\rho}^2} - \frac{2}{3} \frac{g_{\Phi}}{ m_{\Phi}^2})
\frac{f_D m_{D}^2}{m_{D}^2 - m_{K}^2},
\label{apcd0}
\end{eqnarray}
\begin{eqnarray}
A_{PV} (D^{0} \to \bar K^{*0} \gamma)  = 0,
\label{apvd0}
\end{eqnarray}
\begin{eqnarray}
A_{PC} (D^{+}_s \to \rho^{+} \gamma)&  = &4 a_1 | \lambda^{\prime} - \lambda 
\frac{{\tilde g}_V}{3{\sqrt 2}} \frac{g_{\Phi}}{m_{\Phi}^2}|
\frac{f_{D_s} g_{\rho} m_{D_s^*}}{m_{D_s^*}^2 - m_{\rho}^2} 
{\sqrt \frac{m_{D^*_s}}{m_{D_s}}}\nonumber\\
& + & 4 a_1 |C_{VV\Pi}|\frac{g_{\omega}}{3 m_{\omega}^2}   
\frac{f_{D_s} m_{D_s}^2}{m_{D_s}^2 - m_{\pi}^2},
\label{apcds}
\end{eqnarray}
\begin{eqnarray}
A_{PV} (D^{+}_s \to \rho^{+} \gamma)&  = &2 a_1 
\frac{f_{D_s} g_{\rho} }{m_{D_s}^2 - m_{\rho}^2} .
\label{apvds}
\end{eqnarray}

The long distance contribution of the penguin - like operators appears in the 
Cabibbo suppressed ($V_{ud} V_{cd}^*$) decay amplitudes and has the 
$a_2$ Wilson constant as coefficent, 
\begin{eqnarray}
A_{PC} (D^{+} \to \rho^{+} \gamma)&  = &4 a_1 | \lambda^{\prime} - \lambda 
\frac{{\tilde g}_V}{2{\sqrt 2}} (\frac{g_{\rho}}{m_{\rho}^2}
- \frac{g_{\omega}}{3 m_{\omega}^2}) |
\frac{f_D g_{\rho} m_{D^*}}{m_{D^*}^2 - m_{\rho}^2} 
{\sqrt \frac{m_{D^*}}{m_D}}\nonumber\\
& + & 4 a_1 |C_{VV\Pi}|(\frac{g_{\omega}}{3 m_{\omega}^2})   
\frac{f_D m_{D}^2}{m_{D}^2 - m_{\pi}^2}\nonumber\\
& + & 2 a_2  \frac{|V^{D \rho}(0)|}{m_D + m_{\rho}}
[ - \frac{1}{2} \frac{g_{\rho}^2}{m_{\rho}^2}
+ \frac{1}{6} \frac{g_{\omega}^2}{ m_{\omega}^2} + 
\frac{1}{3}\frac{g_{\Phi}^2}{ m_{\Phi}^2}],
\label{apcdpcs}
\end{eqnarray}
\begin{eqnarray}
A_{PV} (D^{+} \to \rho^{+} \gamma)&  = &2 a_1 
\frac{f_D g_{\rho} }{m_{D_s}^2 - m_{\rho}^2} 
\nonumber\\
& + & 2 a_2 \frac{1}{m_D - m_{\rho} }|A_1^{D\rho}(0)|
[ - \frac{1}{2} \frac{g_{\rho}^2}{m_{\rho}^2}
+ \frac{1}{6} \frac{g_{\omega}^2}{ m_{\omega}^2} + 
\frac{1}{3}\frac{g_{\Phi}^2}{ m_{\Phi}^2}]
\label{apvdpcs}
\end{eqnarray}
\begin{eqnarray}
A_{PC} (D^{+}_s \to K^{*+} \gamma)&  = &4 a_1 | \lambda^{\prime} - \lambda 
\frac{{\tilde g}_V}{3{\sqrt 2}} \frac{g_{\Phi}}{m_{\Phi}^2}|
\frac{f_{D_s} g_{K^{*}} m_{D_s^*}}{m_{D_s^*}^2 - m_{K^*}^2} 
{\sqrt \frac{m_{D^*_s}}{m_{D_s}}}\nonumber\\
& + & 2 a_1 |C_{VV\Pi}|(\frac{g_{\omega}}{3 m_{\omega}^2}   
+ \frac{g_{\rho}}{m_{\rho}^2} - \frac{2}{3} \frac{g_{\Phi}}{ m_{\Phi}^2})
\frac{f_{D_s} m_{D_s}^2}{m_{D_s}^2 - m_{K}^2}\nonumber\\
& + & 2a_2 \frac{|V^{D_sK^*}(0)|}{m_{D_s} + m_{K^*}}
[ - \frac{1}{2} \frac{g_{\rho}^2}{m_{\rho}^2}
+ \frac{1}{6} \frac{g_{\omega}^2}{ m_{\omega}^2} + 
\frac{1}{3}\frac{g_{\Phi}^2}{ m_{\Phi}^2}],
\label{apcdscs}
\end{eqnarray}
\begin{eqnarray}
A_{PV} (D^{+}_s \to K^{*+} \gamma)&  = &2 a_1 
\frac{f_{D_s} g_{K^{*}} }{m_{D_s}^2 - m_{K^*}^2}\nonumber\\
 & + & 2 a_2 [ - \frac{1}{2} \frac{g_{\rho}^2}{m_{\rho}^2}
+ \frac{1}{6} \frac{g_{\omega}^2}{ m_{\omega}^2} + 
\frac{1}{3}\frac{g_{\Phi}^2}{ m_{\Phi}^2}] \frac{|A_1^{D_s K^{*}}(0)|}{m_{D_s} -
 m_{K^*}}.
\nonumber\\
\label{apvdscs}
\end{eqnarray}

The Cabibbo suppressed decays of $D^0$ meson involve the contribution from 
the $\eta - \eta^{\prime}$ 
mixing and we take the mixing angle $\theta = -20^0$ \cite{PDG}. We present the decay amplitudes for 
$D^0 \to V^0 \gamma$ ($V^0 =  \rho $, $\omega$, $\Phi$) as
\begin{eqnarray}
A_{PC} (D^{0} \to V^0 \gamma)&  = &4 a_2 b_V^0| \lambda^{\prime} +\lambda 
\frac{{\tilde g}_V}{2{\sqrt 2}} (\frac{g_{\rho}}{m_{\rho}^2}
+ \frac{g_{\omega}}{3 m_{\omega}^2}) |
\frac{f_D g_V m_{D^*}}{m_{D^*}^2 - m_{V}^2} 
{\sqrt \frac{m_{D^*}}{m_D}}\nonumber\\
& + & 4 a_2| C_{VV\Pi} |  f_D m_{D}^2 b^V\nonumber\\
& + & 2 a_2 [ - \frac{1}{2} \frac{g_{\rho}^2}{m_{\rho}^2}
+ \frac{1}{6} \frac{g_{\omega}^2}{ m_{\omega}^2} + 
\frac{1}{3}\frac{g_{\Phi}^2}{ m_{\Phi}^2}]  \frac{|V^{D V^{0}}(0)|}{m_D + m_{V}},
\label{apcd0cs}
\end{eqnarray}
where $b_{\rho}^0 = - 1/{\sqrt 2}$, $b_{\omega}^0 = 1/{\sqrt 2}$ 
and $b_{\Phi}^0 = 1$. The coefficients $b^V$ are obtained 
\begin{equation} 
b^V = \sum_{i=1}^{3} \frac{B^{VP_i}}{m_D^2 - m_{P_i}^2} 
\label{bv}
\end{equation}
where $P_i$ is $\pi$ for $\rho$ and $\omega$ and $K$ for $\Phi$. 
The coefficients $B^{VP_i}$ are given in the Table 1.
\begin{eqnarray}
A_{PV} (D^{0} \to \rho^{0}/ \omega \gamma)&  = &
2  a_2 \frac{|A_1^{D\rho}(0)|}{m_D - m_{\rho}}
[ - \frac{1}{2} \frac{g_{\rho}^2}{m_{\rho}^2}
+ \frac{1}{6} \frac{g_{\omega}^2}{ m_{\omega}^2} + 
\frac{1}{3}\frac{g_{\Phi}^2}{ m_{\Phi}^2}].
\label{apvdp0s}
\end{eqnarray}

For completeness, we give also the decay amplitudes 
of doubly suppressed decays $D^+ \to K^{*+} \gamma$, 
$D^0 \to K^{*0} \gamma$ in which case the amplitudes are 
proportional to $|V_{us} V_{cd}^*|$:
\begin{eqnarray}
A_{PC} (D^{+} \to K^{*+} \gamma)&  = &4 a_1 | \lambda^{\prime} - \lambda 
\frac{{\tilde g}_V}{2{\sqrt 2}} (\frac{g_{\rho}}{m_{\rho}^2}
- \frac{g_{\omega}}{3 m_{\omega}^2}) |
\frac{f_D g_{K^{*}} m_{D^*}}{m_{D^*}^2 - m_{K^*}^2} 
{\sqrt \frac{m_{D^*}}{m_D}}\nonumber\\
& + & 2 a_1 |C_{VV\Pi}|(\frac{g_{\omega}}{3 m_{\omega}^2}   
+ \frac{g_{\rho}}{m_{\rho}^2} - \frac{2}{3} \frac{g_{\Phi}}{ m_{\Phi}^2})
\frac{f_D m_{D}^2}{m_{D}^2 - m_{K}^2},
\label{apcdp}
\end{eqnarray}
\begin{eqnarray}
A_{PV} (D^{+} \to K^{*+} \gamma)&  = &2 a_1 
\frac{f_D g_{K^{*}} }{m_{D}^2 - m_{K^*}^2} ,
\label{apvdp}
\end{eqnarray}
\begin{eqnarray}
A_{PC} (D^{0} \to {K}^{*0} \gamma)&  = &4 a_2 | \lambda^{\prime} + \lambda 
\frac{{\tilde g}_V}{2{\sqrt 2}} (\frac{g_{\rho}}{m_{\rho}^2}
 + \frac{g_{\omega}}{3 m_{\omega}^2}) |
\frac{f_D g_{K^{*}} m_{D^*}}{m_{D^*}^2 - m_{K^*}^2} 
{\sqrt \frac{m_{D^*}}{m_D}}\nonumber\\
& + & 2 a_2 |C_{VV\Pi}|(\frac{g_{\omega}}{3 m_{\omega}^2} - 
\frac{g_{\rho}}{m_{\rho}^2} - \frac{2}{3} \frac{g_{\Phi}}{ m_{\Phi}^2})
\frac{f_D m_{D}^2}{m_{D}^2 - m_{K}^2},
\label{dsapcd0}
\end{eqnarray}
\begin{eqnarray}
A_{PV} (D^{0} \to K^{*0} \gamma)  = 0.
\label{dsapvd0}
\end{eqnarray}

We present numerical results for the parity conserving and parity 
violating amplitudes in Table 2 for the Cabibbo allowed decays 
and for the Cabibbo suppressed decays in Table 3, 
where we denote  ${\cal A}_{PC} = e G_F/ {\sqrt 2} V_{uq_i} V_{cq_j}^* A_{PC}$ 
and ${\cal A}_{PV} = e G_F/ {\sqrt 2} V_{uq_i} V_{cq_j}^* A_{PV}$. 
In our numerical calculation we use $f_{D_s} = 0.240 GeV$  \cite{PDG}, 
$f_{D} = 0.2 GeV$, and we take $|\lambda^{\prime} - 0.32 \lambda| = 
0.052$ $GeV^{-1}$, since we take $\lambda = 0.479$ $GeV^{-1}$ and $g =0.58$. 
We use also $|C_{VV\Pi}| = 0.31$. 
In our estimation we did not analyze the errors arising from the 
experimental data. In the case of $D^* \to D \gamma$ 
decays they can be as large as $100 \%$ \cite{PDG}. 
An additional uncertainty is coming from the couplings taken 
rather far from their mass - shell values, although in $D \to V \gamma$ decays 
we expect that these deviations  are still quite small. 
Unfortunately, the sign of 
$\lambda^{\prime}$, $\lambda$, $C_{VV\Pi}$, and $g_V$, 
can not be determined from the present experimental data and 
therefore, 
we are not able to make concrete predictions for 
the decay rates. However, we notice that the 
penguin - like long distance contribution 
($III$) is quite small comparing to the dominant contributions 
and it amounts a few percent to 
the decay amplitudes. Nevertheless, 
in  the case of neutral charm meson decays 
it can be rather important due to possible cancellation of 
the contributions $I$ and $II$. This contribution is the only source of 
parity violating amplitudes for $D^0 \to \rho^0 (\omega) \gamma$
decay. 

We note that the long  - distance $c \to u \gamma$ contribution 
has a coefficient 
\begin{equation}
C_V = - \frac{1}{2} \frac{g_{\rho}^2(0)}{m_{\rho}^2}
+ \frac{1}{6} \frac{g_{\omega}^2(0)}{m_{\omega }^2}
+ \frac{1}{3} \frac{g_{\Phi}^2(0)}{m_{\Phi}^2}, 
\label{48}
\end{equation}
which we calculate assuming there is no $q^2$ dependence in 
the $g_{V_i}$ values between $m_{V_i}^2$ and $0$. 
Should such a dependence occur, it would obviously affect 
the value of $C_V$ in view of sensitive GIM cancellation involved. 

We come now to the relevance of this $c \to u \gamma$ contribution 
with respect to possible tests 
for new physics in D decays. 

The ratio of decay widths $ R_K = \Gamma (D^+_s \to K^{*+} \gamma)/ 
\Gamma (D^+_s \to \rho^+ \gamma)$ was suggested recently to be used as 
a test of the physics beyond the Standard Model \cite{BFO0}. 
It was noticed that in the case of exact $SU(3)$  symmetry 
$R_K = tan^2 \theta_c$ (up to the phase space factor 
$(q_{K^{*}}/q_{\rho})^3$).  In addition to $SU(3)$ breaking 
effects coming from different masses and couplings, we notice 
that the presence of the penguin - like contribution 
modifies the ratio $R_K$. 
If there is no cancellation among the other contributions to the amplitudes, 
the modification may be several percent only, in the same ballpark 
or even smaller than $SU(3)$ 
- breaking. However, before we gain enough knowledge from experiment on the 
size of the amplitudes, it is rather difficult to 
expect that the sign of new physics can be seen from the 
deviations from this ratio.  In any case, it is instructive to 
note here that a typical figure for the amount of $SU(3)$ breaking 
can be obtained, e. g., by comparing the calculated 
ratio ${\cal A}_{PV} (D_s^+ \to K^{*+} \gamma) /{\cal A}_{PV} 
(D_s^+ \to \rho^+\gamma) $ $ = [g_{K^*} (m_{D_s}^2 - m_{\rho}^2) ]/
[g_{\rho} (m_{D_s}^2 - m_{K^{*}}^2) ] tan \theta_c = 1.24 tan \theta_c$, 
to the symmetry value of $tan \theta_c$.
The ratio $ R_{\rho}  = \Gamma (D^0 \to \rho^{0} \gamma)/ 
\Gamma (D^0\to \bar K^{*0} \gamma)$ offers the same possibility \cite{BGM} 
to look for a deviation from $R_{\rho} = tan^2 \theta_c$. We 
point out  that the same conclusion is valid for $R_{\rho}$ as for $R_K$. \\

{\bf Summary}\\

We have reinvestigated charm meson weak radiative decays 
into light vector mesons, 
systematically including $SU(3)$ symmetry breaking effects using a VDM 
approximation. 
The coupling of charm vector, charm pseudoscalar mesons and photon are 
changed due to this symmetry breaking, as well as Wess-Zumino-Witten 
couplings in the light sector. 

In addition to these known contributions, we have found that 
the long distance penguin - like contribution, proportional to $a_2$ 
Wilson coefficient appears in the charm meson radiative weak Cabibbo 
suppressed decays. The parity conserving and parity violating 
decay amplitudes obtain 
typically a few percent  contribution of the $c \to u \gamma$ 
long distance penguin operator. 

Although this effect is not very large, in the case of neutral 
charm mesons decay it  might play an important role, 
due to possible cancellation of 
the dominant contributions.

At this point, we would like to select and summarize 
those of our results which have smaller 
uncertainties and to compare with previous calculations. 
Among the Cabibbo allowed decays, $D_s^+ \to \rho^+ \gamma$ 
is less affected by interference and we expect it to occur with the significant 
branching ratio $BR (D_s^+ \to \rho^+ \gamma) = (3-5) \times 10^{-4}$. 
Among the Cabibbo suppressed decays, those involving 
less uncertainties are 
$D^+ \to \rho^+ \gamma$ and $D_s^+ \to K^{*+} \gamma$ 
and we expect their occurence with 
$BR(D^+ \to \rho^+ \gamma) = (1.8 - 4.1) \times 10^{-5}$ 
and $BR(D_s^+ \to K^{*+} \gamma) = (2.1 - 3.2) \times 10^{-5}$. 
These results are fairly close to those of Ref. \cite{BGHP}, but of a more 
precise range and present an interesting challenge 
for experiment. We remark that the prediction of Ref. \cite{AK} 
is one order of magnitude smaller. 

Concerning the other decays we calculate, our range of predictions is considerably 
weaker. Thus, $D^0 \to \bar K^{*0} \gamma$ could 
have a branching ratio as high as $3 \times 10^{-4}$, but it also might be 
orders of magnitude smaller as a result of interference between $A^{I}_{PC}$ 
and 
$A^{II}_{PC}$. Its measurement is therefore of great interest. 
Likewise, the decays 
$D^0 \to (\rho^0, \omega^0,  \Phi^0) \gamma$ could have branching ratios 
as large as $2 \times 10^{-5}$, though in case of large negative 
interference among various contributions the branching ratio may be 
reduced by two orders of magnitude. 

A main difference between our results and those of 
Ref. \cite{BGHP} is the amount of parity - violation in various decays. 
These authors treat $D \to V \gamma$ as driven by $D \to V V^{\prime} 
\to  V \gamma$ \cite{GP}. 
However, the D - meson nonleptonic decays are 
rather difficult for any theoretical description \cite{BGHP} 
and the experimental errors on $D \to V V^{\prime}$ are rather large \cite{PDG}.
In any case, this issue will be clarified by the awaited experiments. 

Finally, we note that the suggestion 
that the ratios $ R_K = \Gamma (D^+_s \to K^{*+} \gamma)/ 
\Gamma (D^+_s \to \rho^+ \gamma)$ \cite{BFO0} or 
$ R_{\rho}  = \Gamma (D^0 \to \rho^{0} \gamma)/ 
\Gamma (D^0 \to \bar K^{0} \gamma)$  \cite{BGM} might be useful in a search of 
a signal for physics beyond Standard Model, could be  invalidated 
by the presence of long distance 
penguin - like $c \to u \gamma$ contributions in case of large cancellations 
among various contributions to the amplitudes. 
This effect would affect more the $R_{\rho}$ ratio. \\

{\bf ACKNOWLEDGMENTS}\\

The research of S.F. was supported in part by the Ministry of Science of 
the Republic of Slovenia. 
She thanks for warm hospitality the Physics Department at the Technion. 
The research of P.S. was supported in part by Fund for 
Promotion of Research at the Technion. 
\newpage

\begin{table}[h]
\begin{center}
\begin{tabular}{|c||c|c|c|}\hline
  $B^{VP_i}$ & $\pi$  & $\eta$ & $\eta^{\prime}$ \\ \hline \hline
 $\rho^0$ & $ \frac{1}{3 {\sqrt 2}} \frac{g_{\omega}}{m_{\omega}^2}$ &  
$ - \frac{1}{{\sqrt 2}} c (c - {\sqrt 2} s) \frac{g_{\rho}}{m_{\rho}^2}$ & 
$ - \frac{1}{{\sqrt 2}} s ({\sqrt 2} c + s) 
\frac{g_{\rho}}{m_{\rho}^2}$\\
 $\omega $ & $ \frac{1}{ {\sqrt 2}} \frac{g_{\rho}}{m_{\rho}^2}$ &  
$ - \frac{1}{3{\sqrt 2}} c (c - {\sqrt 2} s) \frac{g_{\omega}}{m_{\omega}^2}$ & 
$ - \frac{1}{3{\sqrt 2}} s ({\sqrt 2} c + s) 
\frac{g_{\omega}}{m_{\omega}^2}$\\
 $\Phi$ & $0$ &  
$ - \frac{{\sqrt 2}}{3} c ({\sqrt 2}c + s) \frac{g_{\Phi}}{m_{\Phi}^2}$ & 
$ - \frac{{\sqrt 2}}{3} s ({\sqrt 2} c - s) 
\frac{g_{\Phi}}{m_{\Phi}^2}$\\
\hline
\end{tabular}
\caption{The $B^{VP_i}$ coefficients defined in relation (19), where 
$s = sin\theta$ $c= cos \theta$, and $\theta$ is the $\eta - \eta^{\prime}$ 
mixing angle.}\label{tab1}
\end{center}
\end{table}

\begin{table}[h]
\begin{center}
\begin{tabular}{|c||c|c|c|}
\hline
$D\to V \gamma$ & $|{\cal A}_{PC}^{I}|$  
& $|{\cal A}_{PC}^{II}|$
& $|{\cal A}_{PV}| $  \\
\hline \hline
$ D^0 \to {\bar K}^{*0} \gamma$ &$ 6.4$ & $6.2$ & $0$  \\
\hline
$ D_s^+ \to \rho^+ \gamma$ &$ 1.4$ & $7.3$ & $7.4$  \\
\hline
\end{tabular}
\caption{ The parity conserving 
%${\cal A}_{PC} = e G_F/ {\sqrt 2} V_{ud} V_{cs}^* A_{PC}$ 
and parity violating 
%${\cal A}_{PV} = e G_F/ {\sqrt 2} V_{ud} V_{cs}^* A_{PV}$
amplitudes for Cabibbo allowed charm meson decays 
in units $10^{-8}$ $GeV^{-1}$. }\label{tab2}
\end{center}
\end{table}

\vskip 1cm
\newpage
\begin{table}[h]
\begin{center}
\begin{tabular}{|c||c|c|c|c|c|}
\hline
$D\to V \gamma$ & $|{\cal A}_{PC}^{I}|$  
& $|{\cal A}_{PC}^{II}| $ 
& $|{\cal A}_{PC}^{III}| $
& $|{\cal A}_{PV}| $ 
& $|{\cal A}_{PV}^{III}| $ \\
\hline \hline
$ D^0 \to \rho^{0} \gamma$ &$ 8.2$ & $10.7$ & $0.2$& $0$ & $0.3$  \\
\hline
$ D^0 \to \omega \gamma$ &$ 7.3$ & $10.7$ & $0.2$& $0$ & $0.3$  \\
\hline
$ D^0 \to \Phi \gamma$ &$ 18.8$ & $13.4$ & $0$& $0$ & $0$  \\
\hline
$ D^+ \to \rho^+ \gamma$ &$ 5.9$ & $13.9$ & $0.2$& $15.9$ & $0.3$  \\
\hline
$ D_s^+ \to K^{*+ }\gamma$ &$ 4.1$ & $23.2$ & $0.2$& $20.8$ & $0.4$  \\
\hline
\hline 
$ D^+ \to K^{*+} \gamma$ &$ 1.6$ & $4.2$ &$ 0$ &  $4.3$ & $ 0$  \\
\hline
$ D^0 \to K^{*0} \gamma$ &$ 3.3$ & $3.2$ &$ 0$ &  $0$ & $ 0$ \\
\hline
\end{tabular}
\caption{ The parity conserving 
%${\cal A}_{PC} = e G_F/ {\sqrt 2} V_{ud} V_{cd}^* A_{PC}$
and parity violating 
%${\cal A}_{PV} = e G_F/ {\sqrt 2} V_{ud} V_{cd}^* A_{PV}$
amplitudes for Cabibbo suppressed 
charm meson decays in units $10^{-9}$ $GeV^{-1}$. 
The last two decays are doubly Cabibbo suppressed. 
%(${\cal A}_{PC} = e G_F/ {\sqrt 2} V_{us} V_{cd}^* A_{PC}$ 
%and ${\cal A}_{PV} = e G_F/ {\sqrt 2} V_{us} V_{cd}^* A_{PV}$).
}\label{tab3}
\end{center}
\end{table}
%\newpage

{\bf Figure Captions}\\

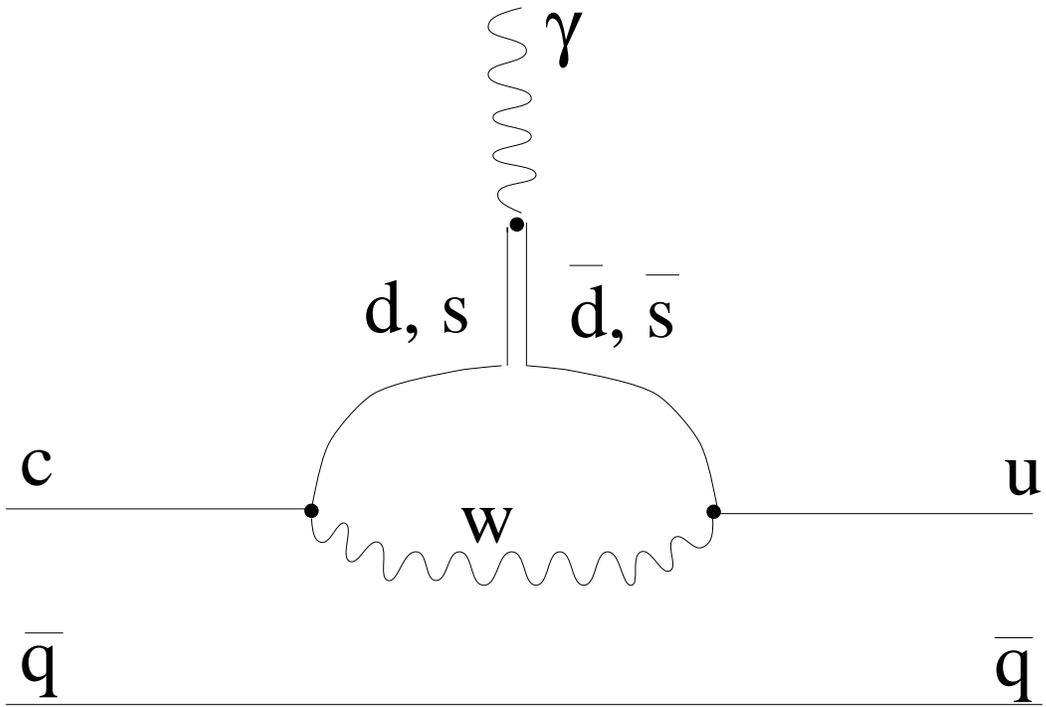
{\bf Fig. 1.} The long distance penguin - like $c \to u \gamma$ transition.\\

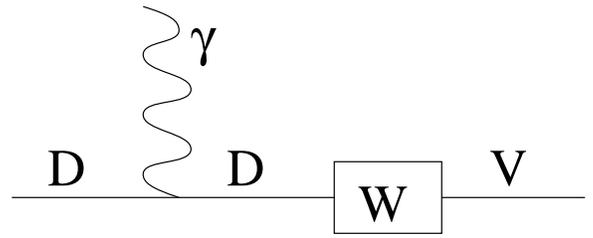
{\bf Fig. 2.} Skeleton graphs for the various contributions to the decay 
amplitudes $D \to V \gamma$. Graph $(1)$ contributes to $A_{PC}^{(I)}$,  
graph $(2)$ to $A_{PC}^{(II)}$ part and graphs $(3a)$, $(3b)$ to $A_{PV}$ 
in Table 2. \\

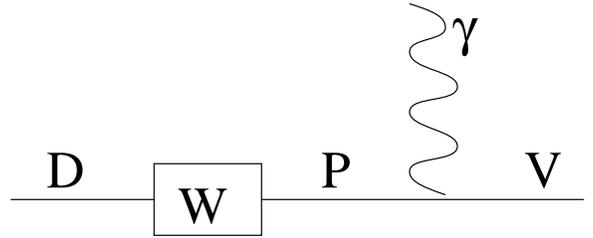
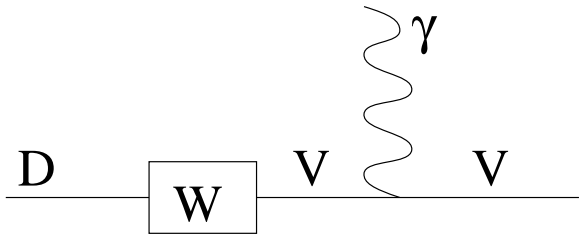
\end{document}